\documentclass[aps,prb,preprint,longbibliography,superscriptaddress]{revtex4-2}
\usepackage{mathrsfs}
\usepackage{amsmath,gensymb}
\usepackage{amsfonts}
\usepackage{amssymb}
\usepackage{amsthm}
\usepackage{graphicx}
\usepackage{natbib}
\usepackage{xcolor}
\usepackage{hyperref}
\usepackage{bm}
\usepackage[caption=false]{subfig}
\usepackage{verbatim,ulem}

\begin{document}
\title{N\'eel proximity effect at antiferromagnet/superconductor interfaces}

\author{G.~A.~Bobkov}
\affiliation{Moscow Institute of Physics and Technology, Dolgoprudny, 141700 Moscow region, Russia}

\author{I.~V.~Bobkova}
\affiliation{Institute of Solid State Physics, Chernogolovka, 142432 Moscow region, Russia}
\affiliation{Moscow Institute of Physics and Technology, Dolgoprudny, 141700 Moscow region, Russia}
\affiliation{National Research University Higher School of Economics, 101000 Moscow, Russia}

\author{A.~M.~Bobkov}
\affiliation{Institute of Solid State Physics, Chernogolovka, 142432 Moscow region, Russia}
\affiliation{Moscow Institute of Physics and Technology, Dolgoprudny, 141700 Moscow region, Russia}

\author{Akashdeep Kamra}
\affiliation{Condensed Matter Physics Center (IFIMAC) and Departamento de F\'{i}sica Te\'{o}rica de la Materia Condensada, Universidad Aut\'{o}noma de Madrid, E-28049 Madrid, Spain}

\begin{abstract}
Spin-splitting induced in a conventional superconductor weakens superconductivity by destroying spin-singlet and creating spin-triplet Cooper pairs. We demonstrate theoretically that such an effect is also caused by an adjacent compensated antiferromagnet, which yields no net spin-splitting. We find that the antiferromagnet produces N\'eel triplet Cooper pairs, whose pairing amplitude oscillates rapidly in space similar to the antiferromagnet's spin. The emergence of these unconventional Cooper pairs reduces the singlet pairs' amplitude, thereby lowering the superconducting critical temperature. We develop a quasiclassical Green's functions description of the system employing a two-sublattice framework. It successfully captures the rapid oscillations in the Cooper pairs' amplitude at the lattice spacing scale as well as their smooth variation on the larger coherence length scale. Employing the theoretical framework thus developed, we investigate this N\'eel proximity effect in a superconductor/antiferromagnet bilayer as a function of interfacial exchange, disorder, and chemical potential, finding rich physics. Our findings also offer insights into experiments which have found a larger than expected suppression of superconductivity by an adjacent antiferromagnet. 
\end{abstract}

\maketitle

\section{Introduction}

Conventional superconductors are formed by spin-singlet Cooper pairs~\cite{Cooper1956,Bardeen1957}. Exposing them to a net spin-splitting, such as via an applied magnetic field or due to interfacial exchange interaction with a ferromagnet (F), causes spin-singlet pairs to be converted into their spin-triplet counterparts~\cite{Maki1964,Maki1969,Bergeret2005,Buzdin2005,Bergeret2018}. This weakens the conventional superconducting state and lowers the critical temperature~\cite{Chandrasekhar1962,Clogston1962,Sarma1963}. On the other hand, since the net magnetization in an antiferromagnet (AF) vanishes, an adjacent superconductor (S) interfaced to the former via a compensated interface is expected to experience no net spin-splitting or reduction in critical temperature~\cite{Kamra2018,Werthammer1966}. Nevertheless, unconventional Andreev reflection and bound states at such S/AF interfaces have been predicted~\cite{Bobkova2005,Andersen2005}. The rich Josephson physics in S/AF/S hybrids has also been investigated, theoretically~\cite{Andersen2006,Bulaevskii2017,Enoksen2013,Rabinovich2019} and experimentally~\cite{Bell2003,Komissinskiy2007,Zhou2019}.

Several experiments have found that AFs lower the critical temperature of an S layer~\cite{Bell2003,Hubener2002,Wu2013,Seeger2021}, despite the no net spin-splitting argument above. In some cases, the effect has been comparable to or even larger than that induced by a ferromagnet layer~\cite{Wu2013}. To understand this, several potential consequences of the AF layer have been considered. First, an AF doubles the spatial period of the lattice due to its antiparallel spins on the two sublattices. This can open a bandgap in the adjacent conductor, which may reduce the normal-state density of states in S and thus, superconductivity~\cite{Buzdin1986,Krivoruchko1996}. Second, it has been shown that an uncompensated interface, which seems to be common in experiments~\cite{Belashchenko2010,Kamra2018,He2010,Manna2014}, to an AF insulator does induce a net spin-splitting~\cite{Kamra2018}, just like a ferromagnet. Furthermore, the interfacial disorder was found to cause spin-flip scattering and reduce superconductivity~\cite{Kamra2018,Abrikosov1961}. While these offer potential mechanisms for affecting the S, they do not account for the phenomena that underlie the previously considered unconventional Andreev reflection~\cite{Bobkova2005,Andersen2005}. Also, the question of how can AFs affect adjacent S more than ferromagnets remains unanswered. Furthermore, a recent Bogoliubov-de Gennes numerical analysis of a hybrid comprising a compensated AF interfaced with an S suggested the interface to be spin-active~\cite{Johnsen2021}. Hence, a key piece of the puzzle in understanding S/AF bilayers appears to be missing.

In this article, we undertake a detailed theoretical investigation of an S/AF bilayer with a compensated interface. For simplicity, we assume the AF to be an insulator. We first analyze this system numerically solving the Bogoliubov-de Gennes equation on a two-dimensional lattice. In this analysis, we find that the N\'eel order of the AF induces spin-triplet correlations in the S. Their amplitude flips sign from one lattice site to the next, just like the N\'eel spin order in the AF. Thus, we call these N\'eel triplet Cooper pairs. The AF with its compensated interface has induced rapidly oscillating triplet correlations, that escape the conventional quasiclassical description of the S since it cannot resolve variations on such short length scales~\cite{Kopnin2001,Belzig1999}. 

In order to adequately account for such rapid oscillations, we develop a two-sublattice quasiclassical Green's functions description of the S. Employing it, we obtain analytic results for the N\'eel pairing amplitude and the critical temperature of the ensuing superconducting state, finding results consistent with our Bogoliubov-de Gennes numerics. Furthermore, the developed framework allows us to address the role of chemical potential in the normal state, disorder, and the strength of interfacial exchange semi-analytically thereby providing valuable insights. 

We find that the N\'eel triplets are formed due to interband pairing within the two-sublattice model. Thus, they are formed when the chemical potential is such that there are two bands around the Fermi surface within the energy $\Delta_0$, the superconducting gap of S without the adjacent AF. The formation of such N\'eel spin-triplets comes at the expense of destroying the spin-singlet correlation which reduces the S critical temperature. On the other hand, disorder destroys this interband pairing and diminishes the proximity effect of the AF on the S. Thus, our investigated mechanism weakens superconductivity strongly for clean systems, yielding an opposite trend  with respect to the role of disorder-mediated spin-flip scattering. Thus, the role of disorder in proximity effect with an AF can be positive or negative on the superconductivity. These competing effects further allow maxima to be feasible in experiments~\cite{Wu2013,Hubener2002,Seeger2021}.

\section{Bogoliubov-de Gennes analysis}\label{sec:BdG}

\begin{figure}[tbh]
	\begin{center}
		\includegraphics[width=85mm]{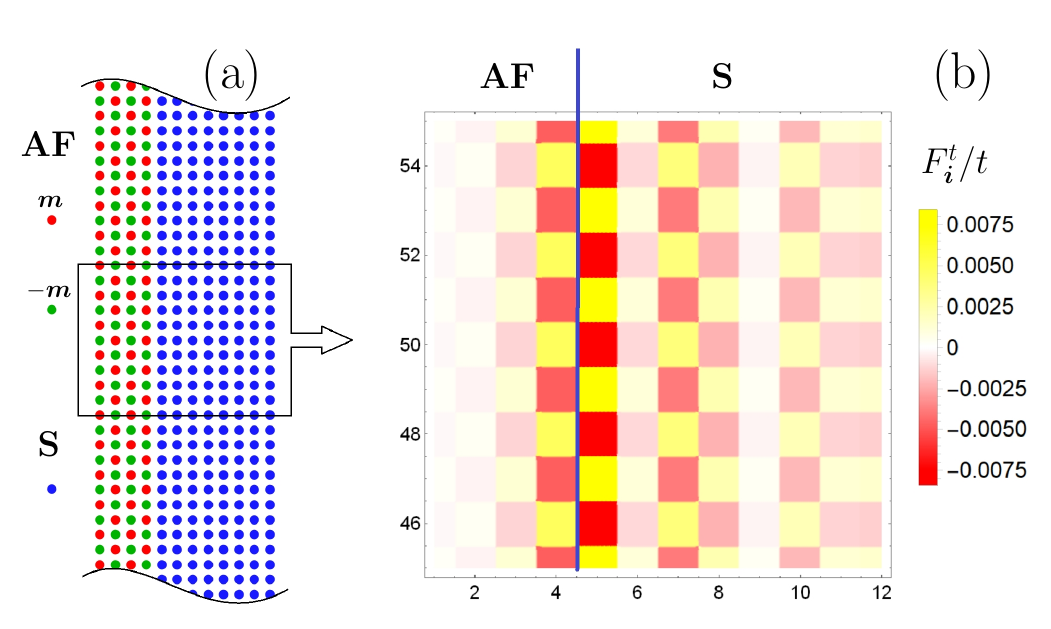}
		\caption{(a) Sketch of the antiferromagnetic insulator interfaced via a compensated interface to the thin superconductor, considered in the framework of a tight-binding Bogoliubov-de Gennes Hamiltonian in Sec.~\ref{sec:BdG}. The system represents a two-dimensional 12 × 100 spatial cluster. Blue points are S sites, red and green points correspond to AF sites with opposite directions of the on-site magnetization $\pm \bm m$. (b) Spatial variation of the triplet correlations amplitude $F_{\bm i}^t$ in the investigated AF/S bilayer. Each colored square codes the value of $F_{\bm i}^t$ at a given site. Only small vertical part of the bilayer is shown, which is marked with a black rectangle in panel (a). An alternating sign of the correlations in S commensurate with the N\'eel order in the AF can be seen along the interfacial direction. The triplet amplitude is normalized to the hopping amplitude, see appendix.} \label{fig:trip}
	\end{center}
\end{figure}

We begin by numerically examining the system of interest - an antiferromagnetic insulator interfaced via a compensated interface to a thin superconductor [Fig.~\ref{fig:trip}(a)]. To this end, we set up a tight-binding Bogoliubov-de Gennes Hamiltonian for the two materials~\cite{Zhu2016} and assume on-site local s-wave correlations in the superconductor, assumed conventional. For simplicity, we consider a two-dimensional system ($12 \times 100$ spatial cluster was considered) and employ periodic boundary conditions along the interfacial direction. The latter emulates an infinite-length interface. The alternating N\'eel spin order in the AF induces a correspondingly oscillating spin-splitting at the interfacial lattice sites in the S via exchange interaction between the AF spins and the S electrons. Describing the Hamiltonian and methodology details in the appendix, we numerically diagonalize the Hamiltonian and evaluate the superconducting state self-consistently.

The spin-triplet correlations amplitude $F_{\bm i}^t$ at each lattice site with the radius-vector $\bm i$ is evaluated by summing the anomalous Green's function over the positive Matsubara frequencies, as detailed in the appendix. Figure \ref{fig:trip}(b) plots the spatially resolved spin-triplet pairing amplitude in the investigated bilayer. A clear imprinting of the AF N\'eel order is seen on the triplet pairing amplitude in the direction parallel to the interface. This perfectly commensurate variation of the triplet correlations is disturbed in the direction perpendicular to the interface by Friedel oscillations~\cite{Zhu2016} and a general lack of out-of-plane momentum conservation due to the interface. A finite $F_{\bm i}^t$ in the first few lattice sites of the AF is due to a small leakage of the electron wavefunctions into the insulating AF. Furthermore, although not explicitly shown, the superconducting critical temperature is found to be reduced substantially by the AF. It is worth noting that if instead we consider a two-sublattice checkerboard ferrimagnet with $m_1 \neq m_2$, this order is also  imprinted on the spin-triplet correlations amplitude: it is a superposition of the perfect N\'eel order presented here and a conventional triplet amplitude, which is homogeneous  along the interface. The same type of the N\'eel triplet order also appears inside a metallic antiferromagnet due to the proximity to a superconductor.

Our numerical analysis clearly demonstrates a large proximity effect of the AF on the S, despite a compensated interface resulting in no net spin-splitting on spatial averaging over the S coherence length scale~\cite{Kamra2018}. It also shows that the interesting physics is taking place on the lattice constant length scale, which is beyond the resolution of the conventional quasiclassical Green's functions description of the superconductor~\cite{Kopnin2001}. Johnsen and coworkers also recently found the compensated interface between AF and S to be spin-active using a similar Bogoliubov-de Gennes numerics~\cite{Johnsen2021}. However, they considered trilayers involving an additional ferromagnet, which prevents a clear understanding of the phenomena taking place at the AF/S interface.

\section{Quasiclassical Green's functions description of a two-sublattice system}

The description of superconductors in terms of quasiclassical Green's functions has proven immensely powerful in understanding all kinds of hybrids involving magnets and superconductors~\cite{Belzig1999,Kopnin2001,Buzdin2005,Bergeret2005,Bergeret2018}. This framework is made tractable by averaging over rapid spatial variations on the Fermi wavelength length scale, which is comparable to or larger than the lattice spacing. Such a procedure adequately captures the properties of the superconducting state while ignoring some small details the underlying normal-metal state.

Motivated and guided by our numerical results based on solving the Bogoliubov-de Gennes equation (Sec.~\ref{sec:BdG}), we wish to develop a quasiclassical Green's function description capable of capturing these effects semi-analytically. In doing so, we notice that the rapid oscillations in the pairing amplitude on the lattice constant scale are merely an expression of the two-sublattice nature of the AF/S system. Thus, these oscillations can be adequately captured by working with a two-sublattice framework and the spatial variations on an individual sublattice are expected to remain slow, as compared to the lattice spacing. This method is directly analogous to capturing the spin order in an antiferromagnet~\cite{Baltz2016}, that shows rapid oscillation between the two sublattices but slow spatial variations within a single sublattice.  

\subsection{Two-sublattice Eilenberger equation}

\begin{figure}[tbh]
	\begin{center}
		\includegraphics[width=85mm]{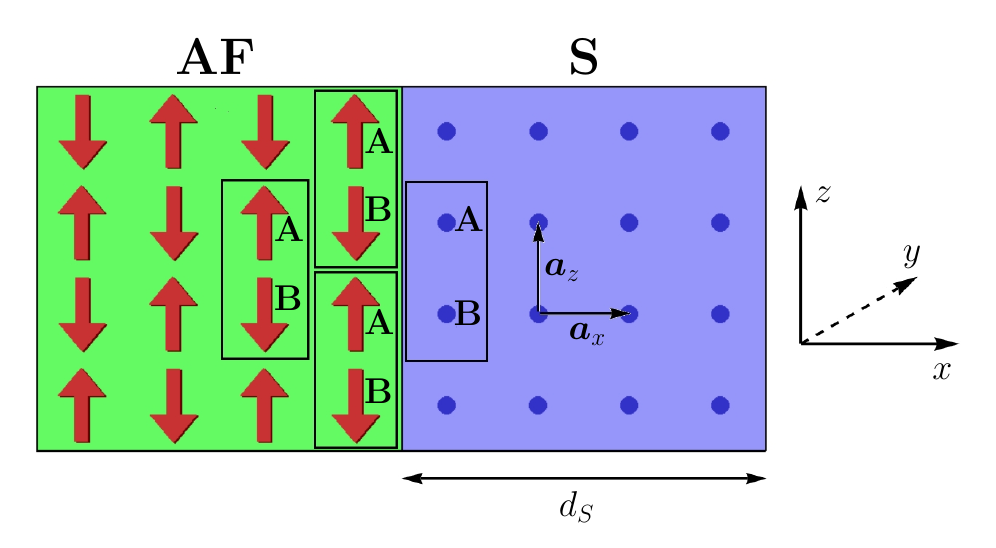}
		\caption{Schematic depiction of the setup under consideration. A N\'eel ordered antiferromagnet (AF) is interfaced via a compensated interface to a superconductor (S). The interface is in the $y-z$ plane and the first superconducting layer is at $i_x=0$. The lattice in both parts of the structure  is divided into two sublattices A and B. The red arrows depict localized spins in the AF. The basis vectors of the original (prior to the introduction of A and B sublattices) lattice in the superconductor are $\bm a_x$, $\bm a_y$, $\bm a_z$.} \label{fig:setup}
	\end{center}
\end{figure}

The unit cell with two sublattices - A and B - is introduced as shown in Fig.~\ref{fig:setup}. In the framework of this two-sublattice approach the unit cells as a whole are marked by radius-vector $\bm i$. Then the staggered magnetism is described by $\bm m_{\bm i,A(B)} = + (-) \bm m_{\bm i}$, where $\bm m_{\bm i}$ is the local magnetic moment at site A of the unit cell with the radius vector $\bm i$ in the AF.  This allows us to consider $\bm m_{\bm i}$  as a slow function of the spatial coordinate. We consider the homogeneously ordered N\'eel state of the AF here, such that $\bm m_{\bm i}$ does not depend on the position $\bm i$. The superconductor S is described by the Hamiltonian: 
\begin{align}
\hat H= &- t \sum \limits_{\langle \bm{i}\bm{j}\nu \bar \nu\rangle ,\sigma} \hat \psi_{\bm{i} \sigma}^{\nu\dagger} \hat \psi_{\bm{j} \sigma}^{\bar \nu} + \sum \limits_{\bm{i},\nu } (\Delta_{\bm{i}}^\nu \hat \psi_{\bm{i}\uparrow}^{\nu\dagger} \hat \psi_{\bm{i}\downarrow}^{\nu\dagger} + H.c.) - \mu \sum \limits_{\bm{i} \nu, \sigma} \hat n_{\bm{i}\sigma}^\nu  \nonumber \\
&+ \sum \limits_{\bm{i} \nu,\alpha \beta} \hat \psi_{\bm{i}\alpha}^{\nu \dagger} (\bm{h}_{\bm{i}}^\nu \bm{\sigma})_{\alpha \beta} \hat \psi_{\bm{i}\beta}^\nu + \sum \limits_{\bm{i}\nu,\sigma} V_i^\nu \hat n_{\bm{i}\sigma}^\nu,
\label{ham_2}
\end{align}
where $\nu=A,B$ is the sublattice index, $\bar \nu = A(B)$ if $\nu=B(A)$ means that the corresponding quantity belongs to the opposite sublattice, $\langle \bm i \bm j \nu \bar \nu\rangle $ means summation over the nearest neighbors, $\hat{\psi}_{\bm i \sigma}^{\nu \dagger}(\hat{\psi}_{\bm i \sigma}^{\nu })$ is the creation (annihilation) operator for an electron with spin $\sigma$ at the sublattice $\nu$ of the unit cell $\bm i$. $t$ parameterizes the hopping between adjacent sites, $\Delta_{\bm{i}}^\nu$ accounts for on-site s-wave pairing, $\mu$ is the electron chemical potential, and $V_i^\nu$ is the local on-site potential that is later employed to capture the effect of impurities and disorder. $\hat n_{\bm i \sigma}^\nu = \hat \psi_{\bm i \sigma}^{\nu\dagger} \hat \psi_{\bm i \sigma}^\nu$ is the particle number operator at the site belonging to sublattice $\nu$ in unit cell $\bm i$. It has been demonstrated that if the AF/S interface is modeled via the interfacial exchange interaction between the localized spins of the antiferromagnet and the spins of conduction electrons in the superconductor, antiferromagnetic order
parameter $\bm m_{i}$ results in the proximity induced exchange field $\bm h_{i} \sim \bm m_i$ on the superconducting side of the interface~\cite{Kamra2018}. Therefore, for the problem under consideration the influence of the antiferromagnetic insulator on the superconductor is described by the  exchange field $\bm h_{\bm i}^{A(B)} = \bm h (-\bm h) \delta_{i_x,0}$, where the $\delta$-symbol means that the exchange field is only nonzero at the AF/S interface sites corresponding to $i_x = 0$ in the superconductor. We assume that the interface is fully compensated, that is the interface exchange field is staggered with zero average value. 

The Matsubara Green's function in the two-sublattice formalism is $8 \times 8$ matrix in the direct product of spin, particle-hole and sublattice spaces. Introducing the two-sublattice Nambu spinor $\check \psi_{\bm i} = (\hat \psi_{{\bm i},\uparrow}^A, \hat \psi_{\bm i,\downarrow}^A, \hat \psi_{\bm i,\uparrow}^B,\hat \psi_{\bm i,\downarrow}^B, \hat \psi_{\bm i,\uparrow}^{A\dagger}, \hat \psi_{\bm i,\downarrow}^{A\dagger}, \hat \psi_{\bm i,\uparrow}^{B\dagger}, \hat \psi_{\bm i,\downarrow}^{B\dagger})^T$ we define the Green's function as follows: 
\begin{eqnarray}
\check G_{\bm i \bm j}(\tau_1, \tau_2) = -\langle T_\tau \check \psi_{\bm i}(\tau_1) \check \psi_{\bm j}^\dagger(\tau_2) \rangle,
\label{Green_Gorkov}
\end{eqnarray}
where $\langle T_\tau ... \rangle$ means  imaginary time-ordered thermal averaging. At first, we consider the clean case corresponding to $V_{\bm i}^\nu = 0$. For the system described by Hamiltonian (\ref{ham_2}) the Heisenberg equation of motion for spinor $\check \psi_{\bm i}$ takes the form: 
\begin{eqnarray}
\frac{d\check \psi_{\bm i}}{d \tau} = [\hat H, \check \psi_{\bm i}] = 
\tau_z \Bigl( t \hat j + \mu - \check \Delta_{\bm i} \sigma_y - \bm h \check {\bm \sigma} \delta_{i_x,0} \rho_z   \Bigr)\check \psi_{\bm i},
\label{psi_motion}
\end{eqnarray}
where $\sigma_k$, $\tau_k$ and $\rho_k$ ($k=x,y,z$) are Pauli matrices in spin, particle-hole and sublattice spaces, respectively. $\bm \sigma = (\sigma_x,\sigma_y,\sigma_z)^T$ is the vector of Pauli matrices in spin space, $\check {\bm \sigma} = \bm \sigma (1+\tau_z)/2 + \bm \sigma^* (1-\tau_z)/2$ is the quasiparticle spin operator and $\check \Delta_{\bm i} = \Delta_{\bm i} \tau_+ + \Delta_{\bm i}^* \tau_-$ with $\tau_\pm  = (\tau_x \pm i \tau_y)/2$. Here we assume that $\Delta_{\bm i}^A = \Delta_{\bm i}^B = \Delta_{\bm i}$, that is the order parameter values are equal for $A$ and $B$ sites of a unit cell. For this reason $\check \Delta_{\bm i}$ is proportional to the unit matrix in the sublattice space. It follows from physical considerations since the change $A \to B$ is equivalent to $\bm h \to -\bm h$ and the order parameter should not depend on odd powers of the exchange field for the problem under consideration. Also this physical assumption is confirmed by subsequent self-consistent calculations. In general, for a case of non-equivalent sublattices matrix $\check \Delta_{\bm i}$ is also diagonal in the sublattice space, but the diagonal components can be not equal $\check \Delta_{\bm i}^A \neq \check \Delta_{\bm i}^B$.    
\begin{eqnarray}
\hat j \check \psi_{\bm i} = \rho_+ \sum \limits_{\bm a} \check \psi_{\bm i+\bm a -\bm a_z } + \rho_- \sum \limits_{\bm a} \check \psi_{\bm i+\bm a +\bm a_z }.
\label{j}
\end{eqnarray}
Here $\bm a \in \{\pm\bm a_x, \pm\bm a_y, \pm\bm a_z \}$, see Fig.~\ref{fig:setup} for definition of these vectors. $\rho_\pm  = (\rho_x \pm i \rho_y)/2$. The Green's function Eq.~(\ref{Green_Gorkov}) obeys equation:
\begin{eqnarray}
\frac{d \check G_{\bm i \bm j}}{d \tau_1} = -\delta (\tau_1-\tau_2)\delta_{\bm i \bm j} - \langle T_\tau \frac{d \check \psi_{\bm i}(\tau_1)}{d \tau_1} \check \psi_{\bm j}^\dagger(\tau_2)\rangle .
\label{Green_gorkov_eq}
\end{eqnarray}
Substituting Eq.~(\ref{psi_motion}) into Eq.~(\ref{Green_gorkov_eq}) and expressing the Green's function in terms of the Matsubara frequencies $\omega_m = \pi T(2m+1)$ we obtain:
\begin{align}
&G_{\bm i,l}^{-1} \check G_{\bm i \bm j}(\omega_m) = \delta_{\bm i \bm j}, \label{gorkov_eq_ml} \\
&G_{\bm i,l}^{-1} = \tau_z \left( t \hat j + \mu - \check \Delta_{\bm i} \i \sigma_y - \bm h \check {\bm \sigma} \delta_{i_x,0} \rho_z  \right) + i \omega_m .
\label{G_i}
\end{align}
Analogously one obtains the equation, where the operator $G^{-1}$ acts on the Green's function from the right:
\begin{align}
&\check G_{\bm i \bm j}(\omega_m)G_{\bm j,r}^{-1} = \delta_{\bm i \bm j}, \label{gorkov_eq_mr} \\
&G_{\bm j,r}^{-1} = \left( t \hat j + \mu + \check \Delta_{\bm i} \i \sigma_y - \bm h \check {\bm \sigma} \delta_{i_x,0} \rho_z  \right) \tau_z  + i \omega_m .
\label{G_j}
\end{align}
where the operator $\hat j$ acts on the Green's function from the right as follows
\begin{equation}
\check G_{\bm i \bm j} \hat j =  \sum \limits_{\bm a} \check G_{\bm i,\bm j+\bm a-\bm a_z}\rho_- +  \sum \limits_{\bm a} \check G_{\bm i,\bm j+\bm a+\bm a_z}\rho_+.
\label{G_j_right}
\end{equation}
Further we consider the Green's function in the mixed representation:
\begin{eqnarray}
\check G(\bm R, \bm p) = F(\check G_{\bm i \bm j}) = \int d^3 r e^{-i \bm p(\bm i - \bm j)}\check G_{\bm i \bm j},
\label{mixed}
\end{eqnarray}
where $\bm R=(\bm i+\bm j)/2$ and the integration is over $\bm i - \bm j$. The term $\hat j \check G_{ij}$ in the mixed representation takes the form:
\begin{eqnarray}
F(\hat j \check G_{ij}) = t e^{ip_z a_z} \rho_+  \sum \limits_{\bm a} e^{-i \bm p \bm a} \check G(\bm R+\frac{\bm a}{2}-\frac{\bm a_z}{2}) + t e^{-ip_z a_z} \rho_-  \sum \limits_{\bm a} e^{-i \bm p \bm a} \check G(\bm R+\frac{\bm a}{2}+\frac{\bm a_z}{2}), ~~~~~~
\label{mixed_hopping}
\end{eqnarray}
$a_i$ is the absolute value of $\bm a_i$ for $i=x,y,z$. Now we define the following  transformed Green's function:
\begin{eqnarray}
\check {\tilde G}(\bm R, \bm p) = 
\left(
\begin{array}{cc}
1 & 0 \\
0 & -i\sigma_y
\end{array}
\right)_\tau \rho_x e^{\frac{-\displaystyle ip_z a_z \rho_z}{\displaystyle 2}} \check G(\bm R, \bm p)  e^{ \frac{\displaystyle ip_z a_z \rho_z}{\displaystyle 2}}
\left(
\begin{array}{cc}
1 & 0 \\
0 & -i\sigma_y
\end{array}
\right)_\tau ,
\label{unitary}
\end{eqnarray}
where subscript $\tau$ means that the explicit matrix structure corresponds to the particle-hole space. Then taking into account Eq.~(\ref{mixed_hopping}) from Eq.~(\ref{gorkov_eq_ml}) we obtain:
\begin{eqnarray}
\left[ i \omega_m \tau_z + \mu + \tau_z \check \Delta(\bm R) - \bm h (\bm R) \bm \sigma  \tau_z \rho_z \right]\rho_x \check {\tilde G}(\bm R, \bm p) + t \frac{\rho_0+\rho_z}{2}\sum \limits_{\bm a}e^{-i \bm p \bm a} \check {\tilde G}(\bm R+ \frac{\bm a}{2}-\frac{\bm a_z}{2},\bm p)   \nonumber \\
+ t \frac{\rho_0-\rho_z}{2}\sum \limits_{\bm a}e^{-i \bm p \bm a} \check {\tilde G}(\bm R+ \frac{\bm a}{2}+\frac{\bm a_z}{2},\bm p)  = 1 ,~~
\label{eilenberger_left} 
\end{eqnarray}
where $\rho_0$ is the unit matrix in the sublattice space. The dependence of the Green's function on $\bm R$ is assumed to be slow, therefore from now we can consider $\bm R$ as a continuous variable. We also generalized our consideration to the case of an arbitrary spatial profile of the exchange field $\bm h(\bm R)$.

Analogously, from Eq.~(\ref{gorkov_eq_mr}) it follows:
\begin{eqnarray}
 \check {\tilde G}(\bm R, \bm p) \left[ i \omega_m \tau_z + \mu + \tau_z \check \Delta(\bm R) - \bm h(\bm R) \bm \sigma \tau_z \rho_z \right]\rho_x  + t \sum \limits_{\bm a}e^{i \bm p \bm a} \check {\tilde G}(\bm R+ \frac{\bm a}{2}+\frac{\bm a_z}{2},\bm p) \frac{\rho_0+\rho_z}{2}   \nonumber \\
+ t \sum \limits_{\bm a}e^{i \bm p \bm a} \check {\tilde G}(\bm R+ \frac{\bm a}{2}-\frac{\bm a_z}{2},\bm p) \frac{\rho_0-\rho_z}{2}  = 1 .~~
\label{eilenberger_right} 
\end{eqnarray}
Subtracting Eqs.~(\ref{eilenberger_left}) and (\ref{eilenberger_right}) we obtain:
\begin{eqnarray}
\left[ \left(i \omega_m \tau_z + \mu + \tau_z \check \Delta(\bm R) - \bm h(\bm R) \bm \sigma \tau_z \rho_z \right)\rho_x , \check {\tilde G}(\bm R, \bm p) \right] +\check A = 0,
\label{eilenberger_sub}
\end{eqnarray}
where,
\begin{align}
\check A = & t \frac{\displaystyle \rho_0+\rho_z}{\displaystyle 2}\sum \limits_{\bm a}e^{-i \bm p \bm a} \check {\tilde G}(\bm R+ \frac{\bm a}{2}-\frac{\bm a_z}{2},\bm p) + t \frac{\displaystyle \rho_0-\rho_z}{\displaystyle 2}\sum \limits_{\bm a}e^{-i \bm p \bm a} \check {\tilde G}(\bm R+ \frac{\bm a}{2}+\frac{\bm a_z}{2},\bm p)  \nonumber \\
   & - t \sum \limits_{\bm a}e^{i \bm p \bm a} \check {\tilde G}(\bm R+ \frac{\bm a}{2}+\frac{\bm a_z}{2},\bm p) \frac{\displaystyle \rho_0+\rho_z}{\displaystyle 2} - t \sum \limits_{\bm a}e^{i \bm p \bm a} \check {\tilde G}(\bm R+ \frac{\bm a}{2}-\frac{\bm a_z}{2},\bm p) \frac{\displaystyle \rho_0-\rho_z}{\displaystyle 2} .
\label{eilenberger_sub_A} 
\end{align}
Now we introduce quasiclassical $\xi$-integrated Green's function:
\begin{eqnarray}
\check g(\bm R, \bm p_F) = -\frac{1}{i \pi} \int \check {\tilde G}(\bm R, \bm p)d\xi,
\label{quasi_green} 
\end{eqnarray}
where $\xi(\bm p) = -2t (\cos p_x a_x+\cos p_y a_y+\cos p_z a_z)$ is the normal state electron dispersion counted from the Fermi energy.   In general, in the framework of the two-sublattice formalism the Brillouin Zone (BZ) is reduced along the $p_z$ direction, that is $p_z \in [-\pi/2a_z, \pi/2a_z]$. This results in appearance of two normal state quasiparticle dispersion branches $\xi_\pm = \mp 2t(\cos p_x a_x+\cos p_y a_y+\cos p_z a_z)-\mu$ in the reduced BZ. Only one of them has a Fermi surface in the first BZ. Let it be $\xi_+ $, then the Fermi momentum $\bm p_F$ is determined by the equation $\xi_+(\bm p_F) = 0$. The quasiclassical Green's function (\ref{quasi_green}) depends only on the direction of the momentum on the Fermi surface, as usual. Please note that in the framework of the developed formalism $\mu$ is assumed to be small with respect to $t$ and, therefore, can be neglected in $\xi_+(\bm p)$, which is reduced to $\xi(\bm p)$ in this case. On the other hand, the Fermi energy (as measured from the bottom of the band) and momentum are large such that the usual conditions for the validity of the quasiclassical theory remain valid. Moreover, if $\mu$ is of the order of $t$, the quasiclassical formalism developed here fails to work. Technically it is because $\mu \sim t $ still enters Eq.~(\ref{eilenberger_sub}) in contrast to the well-known one-sublattice formalism, and the quasiclassical formalism does not allow such a high-energy term. Physically it means suppression of all the correlations between electrons belonging to different normal state dispersion branches and Umklapp scattering processes\cite{Cheng2014,Takei2014,Baltz2016}, see Sec.~\ref{sec:relate}  for further details. In the framework of the quasiclassical approximation $\check {\tilde G}(\bm R + \bm a/2 \pm \bm a_z/2,\bm p) \approx \check {\tilde G}(\bm R, \bm p) + (\bm a \pm \bm a_z) \bm \nabla \check {\tilde G}(\bm R, \bm p)/2$ and we obtain
\begin{eqnarray}
-\frac{1}{i \pi} \int \check A d\xi = it \sum \limits_{\bm a} \sin \bm p_F \bm a (\bm \nabla \check g(\bm R, \bm p_F) \bm a) = i \bm v_F \bm \nabla \check g(\bm R, \bm p_F),
\label{quasi_A} 
\end{eqnarray}
where $\bm v_F = d\xi/d \bm p|_{\bm p = \bm p_F} = 2t(\bm a_x \sin[p_x a_x] + \bm a_y \sin[p_y a_y]+ \bm a_z \sin [p_z a_z])$. After the $\xi$-integration the resulting Eilenberger equation for the quasiclassical Green's function takes the form:
\begin{align}
\left[ \left(i \omega_m \tau_z + \mu + \tau_z \check \Delta(\bm R) - \bm h (\bm R) \bm \sigma \tau_z \rho_z \right)\rho_x , \check g(\bm R,\bm p_F) \right] + i \bm v_F \bm \nabla \check g(\bm R,\bm p_F) = 0 .
\label{eilenberger_ballistic} 
\end{align}
Eq.~(\ref{eilenberger_ballistic}) should be supplemented by the normalization condition
\begin{eqnarray}
\check g^2(\bm R, \bm p_F) = 1 ,
\label{norm}
\end{eqnarray}
which is typical for the quasiclassical theory. In order to verify its validity in our case one can multiply Eq.~(\ref{eilenberger_ballistic}) by $\check g(\bm R, \bm p_F)$ from the left, then from the right and add the resulting equations. This procedure leads to the conclusion that $\check g^2(\bm R, \bm p_F)$ obeys the same Eq.~(\ref{eilenberger_ballistic}) and, therefore, $\check g^2(\bm R, \bm p_F) = 1$ is a solution of this equation for arbitrary spatial profiles of $\check \Delta(\bm R)$ and $\bm h(\bm R)$. In particular, one can assume that $\check \Delta(\bm R)$ and $\bm h(\bm R)$ evolve smoothly to zero values, that is the system transforms to a normal metal. Given that Eq.~(\ref{norm}) holds in the normal metal one can conclude that $\check g(\bm R, \bm p_F)$ obeying Eqs.~(\ref{eilenberger_ballistic}) and (\ref{norm}) is a true solution of the problem under consideration. Below we directly show that Eq.~(\ref{norm}) is valid in the limit of the normal metal $\check \Delta(\bm R) \to 0$ and $\bm h(\bm R) \to 0$.  

Equations \eqref{eilenberger_ballistic} and \eqref{norm} are the desired two-sublattice Eilenberger equation together with the normalization condition on the quasiclassical Green's function matrix. These constitute the main result of this subsection. The sublattice degree of freedom adds the $\rho_{x,y,z}$ Pauli matrices to the framework. It is worth noting that Eq.~(\ref{eilenberger_ballistic}) can also be employed for treating interfaces with ferrimagnets, when $\bm h^A \neq -\bm h^B$. In this case $\bm h(\bm R)\rho_z$ in Eq.~(\ref{eilenberger_ballistic}) is changed by a diagonal matrix in the sublattice space $\hat {\bm h}(\bm R) = \bm h^A(\bm R)(1+\rho_z)/2 + \bm h^B(\bm R)(1-\rho_z)/2$. In the limit $\bm h^A = \bm h^B$ it corresponds to the ferromagnetic case. Also, Eq.~(\ref{eilenberger_ballistic}) adequately describes an antiferromagnetic metal if the exchange field is small as compared to the Fermi energy $\bm h \ll \varepsilon_F$.

\subsection{Inclusion of disorder}

So far, we have considered a homogeneous and clean superconductor. A key strength of the quasiclassical framework is that it can account for several nonidealities, such as impurity scattering or disorder. These are especially important in modeling and understanding experiments which employ superconductors with a varying degree of disorder.

Considering impurity potential in the superconductor in the framework of the Born approximation~\cite{Kopnin2001}, it can be obtained that Eq.~(\ref{eilenberger_sub}) is modified as follows:
\begin{eqnarray}
\Bigl[ \Bigl(i \omega_m \tau_z + \mu + \tau_z \check \Delta(\bm R) - \bm h(\bm R) \bm \sigma \tau_z \rho_z \Bigr)\rho_x - \check \Sigma_{\mathrm{imp}}(\bm R) , \check {\tilde G}(\bm R, \bm p) \Bigr] + \check A = 0 , ~~~~~~~
\label{Eilenberger_impurities} 
\end{eqnarray}
where the impurity self-energy $\check \Sigma_{\mathrm{imp}}$ takes the form:
\begin{eqnarray}
\check \Sigma_{\mathrm{imp}}  = \frac{1}{\pi N_F \tau}\int \frac{d^3p}{(2\pi)^3}[\rho_+ \check {\tilde G}(\bm R, \bm p) \rho_+ + 
\rho_- \check {\tilde G}(\bm R, \bm p) \rho_-] \nonumber \\ 
 = \frac{1}{\pi N_F \tau}\int d\xi \langle N_F(\bm p_F) [\rho_+\check {\tilde G}(\bm R, \bm p)\rho_+ + \rho_-\check {\tilde G}(\bm R, \bm p)\rho_-] \rangle_{\bm p_F},~~~~~~
\label{impurity_self_0} 
\end{eqnarray}
where $\langle ... \rangle_{\bm p_F}$ means Fermi-surface averaging, $N_F(\bm p_F)$ is the momentum direction-dependent density of states (DOS) at the Fermi surface. $N_F$ is the momentum - averaged DOS at the Fermi-surface. $\tau$ is the quasiparticle mean free time, which is connected to the on-site impurity potential $V_i$ as $\langle V_i V_j \rangle  = (1/\pi N_F \tau)\delta_{ij}$. In this case one can express the impurity self-energy Eq.~(\ref{impurity_self_0}) via the quasiclassical Green's function:
\begin{eqnarray}
\check \Sigma_{\mathrm{imp}}(\bm R) = -\frac{i}{\tau}[\rho_+ \check {\bar g}(\bm R) \rho_+ + \rho_- \check {\bar g}(\bm R) \rho_-],
\label{impurity_self} 
\end{eqnarray}
where $\check {\bar g}(\bm R)=\langle \frac{N_F(\bm p_F)}{N_F}  g(\bm R, \bm p_F)\rangle_{\bm p_F}$. Then Eq.~(\ref{eilenberger_ballistic}) in the presence of impurities takes the form:
\begin{eqnarray}
\Bigl[ \Bigl(i \omega_m \tau_z + \mu + \tau_z \check \Delta(\bm R) - \bm h(\bm R) \bm \sigma \tau_z \rho_z \Bigr)\rho_x - \check \Sigma_{\mathrm{imp}}(\bm R) , \check g(\bm R, \bm p_F) \Bigr] + i \bm v_F \bm \nabla \check g(\bm R, \bm p_F) = 0 .~~~~
\label{eilenberger_imp} 
\end{eqnarray}

\subsection{Green's function of a homogeneous normal metal at an arbitrary disorder}
\begin{figure}[tb]
	\centering
		\includegraphics[width=85mm]{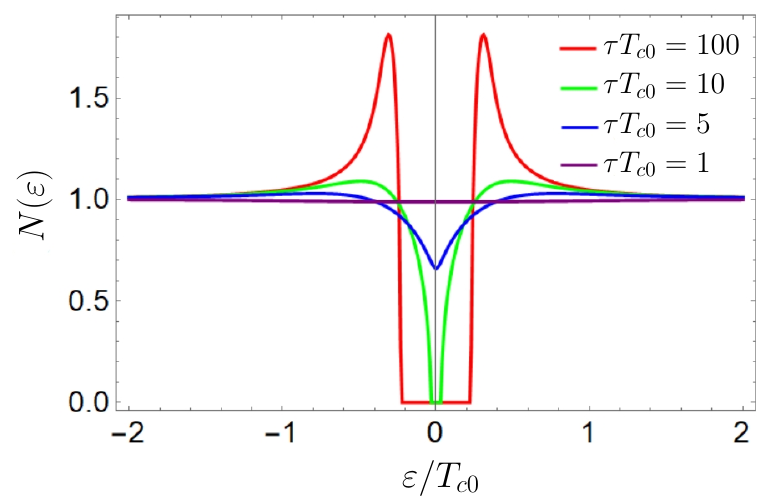}
	\caption{Normal state LDOS for different values of impurity scattering time $\tau$, which is measured in units of the inverse bulk superconducting critical temperature $T_{c0}^{-1}$. $\varepsilon$ is measured in units of $T_{c0}$. We consider $\mu = 0$ and $h=0.3 T_{c0}$ in this figure.}
	\label{ldos_normal} 
\end{figure}

Now let's find the quasiclassical Green's function in the spatially homogeneous normal metal in the presence of spatially homogeneous, but staggered, exchange field $\bm h$ as produced by a compensated antiferromagnet interface. Then the Green's function $\check {\tilde G}(\bm R,\bm p)$ does not depend on $\bm R$ and Eq.~(\ref{eilenberger_left}) is reduced to:
\begin{eqnarray}
\Bigl[\Bigl( i \omega_m \tau_z + \mu  - \bm h \bm \sigma \tau_z \rho_z \Bigr)\rho_x -\xi(\bm p) - \check \Sigma_{\mathrm{imp}} \Bigr] \check {\tilde G}(\bm p) = 1 ,~~~~
\label{G_normal} 
\end{eqnarray}
where we have also added the impurity self-energy. In the clean limit $\tau \to \infty$, i.e., if we disregard impurity scattering, Eq.~(\ref{G_normal}) can be easily solved resulting in the following answer for the $\xi$-integrated Green's function Eq.~(\ref{quasi_green}):
\begin{eqnarray}
\check g = \tau_z \frac{\bm h\bm \sigma \rho_y + (\omega_m - i \tau_z \mu)\rho_x}{\sqrt{h^2+(\omega_m-i \mu \tau_z)^2}} .
\label{g_normal_clean}
\end{eqnarray}
It is seen that the normalization condition Eq.~(\ref{norm}) immediately follows from Eq.~(\ref{g_normal_clean}). It is important that the N\'eel structure of the exchange field radically changes the well-known answer for the quasiclassical Green's function in the normal metal $\check g_N = 1$, which is valid for the case of ferromagnet-type exchange field $\bm h$. This is primarily because the staggered spin-splitting due to the antiferromagnet opens a gap in the normal-state local density of states (LDOS) of the superconductor: 
\begin{eqnarray}
N(\varepsilon) = \frac{1}{4}{\rm Tr}_8\Bigl[\frac{1+\tau_z}{2}\rho_x {\rm Re}[\check g(i \omega_m \to \varepsilon+i \delta)]\Bigr]
\end{eqnarray}
at $\varepsilon = -\mu $. The LDOS at $\mu = 0$ and in the absence of impurities is shown in Fig.~\ref{ldos_normal} by the red line. The LDOS at finite $\mu$ is obtained by shifting the corresponding plots along the energy axis $N_{\mu \neq 0}(\varepsilon) = N_{\mu=0}(\varepsilon + \mu)$. 

In the presence of impurities solution of Eq.~(\ref{G_normal}) can not be written explicitly, but the quasiclassical Green's function still has the off-diagonal structure in the sublattice space, that is
\begin{eqnarray}
\check g = 
\left(
\begin{array}{cc}
0 & \check g^{AB} \\
\check g^{BA} & 0
\end{array}
\right),
\label{g_matrix_imp}
\end{eqnarray}
where
\begin{eqnarray}
\check g^{AB} = \sqrt{[\check \beta + \frac{i}{\tau}\check g^{BA}]^{-1}[\check \alpha + \frac{i}{\tau}\check g^{AB}]}, \nonumber \\
\check g^{BA} = \sqrt{[\check \alpha + \frac{i}{\tau}\check g^{AB}]^{-1}[\check \beta + \frac{i}{\tau}\check g^{BA}]}
\label{g_offdiagonal}
\end{eqnarray}
with
\begin{eqnarray}
\check \alpha (\check \beta) = \mp \bm h \bm \sigma \tau_z + i (\omega_m \tau_z -i\mu).
\label{alpha_beta}
\end{eqnarray}
From Eqs.~(\ref{g_matrix_imp}) and (\ref{g_offdiagonal}) it follows that in the presence of impurities the normalization condition Eq.~(\ref{norm}) also holds. 

Irrespective of the impurity strength the general answer for normal quasiclassical Green's function takes the form:
\begin{eqnarray}
\check g = \left(
\begin{array}{cc}
\hat g_N & 0 \\
0 & \hat {\tilde g}_N
\end{array}
\right)_{\tau},
\label{gn_general}
\end{eqnarray}
where the electron and hole components of the Green's function take the following structure in the sublattice space:
\begin{eqnarray}
\hat g_N = A \rho_x + B \rho_y \bm n_h \bm \sigma, \nonumber \\
\hat {\tilde g}_N = \tilde A \rho_x + \tilde B \rho_y \bm n_h \bm \sigma ,
\label{gn_components}
\end{eqnarray}
$\bm n_h = \bm h/h$. The difference between two sublattices is contained in the second terms in Eq.~(\ref{gn_components}). It is seen that it is directly determined by the presence of the staggered exchange field. The normal state LDOS at finite impurity concentration is presented in Fig.~\ref{ldos_normal}. It is seen that the gap in the LDOS is graduallly suppressed with impurity strength.

\subsection{Relating single-sublattice and two-sublattice pictures}\label{sec:relate}

\begin{figure}[tb]
	\centering
	\includegraphics[width=85mm]{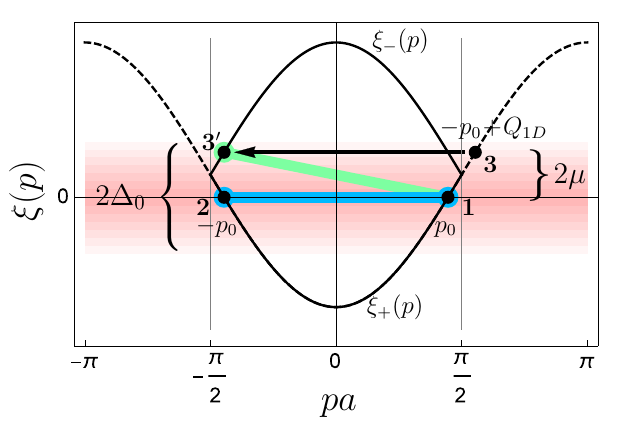}
	\caption{Electron dispersion of the normal-state S in the reduced Brillouin zone (BZ) $pa \in [-\pi/2,\pi/2]$ considering a 1D system with two sites in the unit cell $\xi_{\pm}(p) = \mp 2t \cos pa - \mu$. The reciprocal lattice vector due to the periodicity enforced by the AF is $Q_{1D} = \pi/a$. The spectrum branches are doubled in the BZ due to the reduction of the BZ volume. The blue line indicates ordinary pairing between $(p_0, \xi_1 = 0)$(1) and $(- p_0,\xi_2 = 0)$(2) electrons corresponding to the zero total pair momentum. The green line indicates N\'eel pairing between $p_0$(1) and $- p_0 + Q_{1D}$ (3) corresponding to the total pair momentum $ Q_{1D}$. From the point of view of the first BZ it is an interband pairing between electrons (1) and (3'). Taking into account that $p_0$ is defined from the condition $-2 t \cos p_0 a - \mu = 0$ one immediately obtains that $\xi_1 - \xi_{3'} = 2\mu$. That is, the energy difference between (1) and (3') electrons grows with $\mu$ thus reducing the efficiency of pairing. The antiferromagnetic gap opening as discussed in Fig.~\ref{ldos_normal} has been disregarded in the present simplified figure. $\Delta_0$ is the zero-temperature gap of the bulk S.}
	\label{fig:pairing} 
\end{figure}

We now discuss a physical picture for the N\'eel Cooper pairs. The essential physics is captured conveniently within a one-dimensional (1D) model and we consider that here for clarity~\cite{Simensen2020}. In this model, a 1D AFM is interfaced to a 1D superconductor running along the AFM. Therefore, the electron wavevector bears only one component which is along the interface. 

Now, if we disregard the AFM, the normal-state electronic dispersion of S can be depicted as in Fig.~\ref{fig:pairing} with a BZ $pa \in [-\pi,\pi]$, where $a$ is the lattice constant. Within this single-sublattice dispersion for the normal-state of S, the AF N\'eel order imprinted on the S via a staggered spin-splitting field causes scattering between electronic states that differ by the wavenumber $Q_{1D} = \pi/a$, which is the reciprocal state unit vector for the AF. This has sometimes been termed Umklapp scattering in the literature~\cite{Cheng2014,Takei2014,Baltz2016}. However, we should keep in mind that the momentum change is the reciprocal unit vector of the AF, and not the S. Thus, the AF converts conventional spin-singlet pairing between $+p$ and $-p$ electronic states into N\'eel spin-triplet pairing between, for example, $+p$ and $-p + Q_{1D}$ (see Fig.~\ref{fig:pairing}). Such a pairing oscillates rapidly in space similar to the N\'eel order with the wavenumber $Q_{1D}$.

In the discussion above, we have disregarded the gap opening (see Fig.~\ref{ldos_normal}) caused by the imposition of staggered spin-splitting on the S. This gap opening seems natural when we recognize that the N\'eel ordered AF reduces the periodicity of the adjacent S by imposing a staggered spin-splitting on it and reducing the BZ to $pa \in [-\pi/2,\pi/2]$. Thus, within this adequately rigorous two-sublattice picture, the gap opening is natural and happens at the BZ boundary~\cite{Simensen2020,Krivoruchko1996}. We continue to ignore it in the ongoing discussion and in Fig.~\ref{fig:pairing}, returning to the gap in understanding further physical phenomena below. Furthermore, within this two-sublattice picture, we now have two bands in the electronic dispersion. What appeared as pairing between $+p_0$ and $-p_0 + Q_{1D}$ states in the single-sublattice picture is actually pairing between the $+p_0$ state from one band with the $-p_0$ state from the other band, as depicted in Fig.~\ref{fig:pairing}.

We also see that the N\'eel or interband pairing is taking place between two states which generally have different energies. They becomes degenerate only under the special condition when the chemical potential $\mu$ is zero so that we have perfect half filling in the system, and when we disregard the gap opening due to the staggered spin-splitting. As Cooper pairing takes place between electrons within an energy $\Delta_0$ (the superconducting gap of the bulk S) from the Fermi surface, the strength of interband pairing depends on the LDOS in the normal-state. This understanding will be employed to qualitatively understand several of our results below. 

In principle, our two-sublattice formalism can be reformulated in terms of the two-band picture. The resulting equations would be to some extent similar to the quasiclassical formalism developed for two-band superconductors, where superconductivity and
spin density waves coexist~\cite{Moor2011,Dzero2015}. Further, it is interesting to compare the suppression of superconductivity by a compensated AF studied herein to the destruction of superconductivity by magnetic impurities~\cite{Abrikosov1961}. In the case under investigation, the conversion of singlet Cooper pairs into N\'eel triplets, the latter having zero amplitude on spatial averaging, lowers the critical temperature. On the other hand, unordered magnetic impurities create triplet Cooper pairs with random spin polarization axes thereby suppressing singlets and superconductivity.

\section{Critical temperature of the superconductor/antiferromagnet bilayer}

We now employ the quasiclassical formalism developed in the previous section to study the proximity effect and the superconducting critical temperature in the AF/S bilayer. The S layer is assumed to be thin with respect to the coherence length i.e., $d_S \ll \xi_S$. In this case the Green's function $\check g$ is approximately constant in the superconductor except for the trajectories nearly parallel to the AF/S interface~\cite{Eschrig2015,Eschrig2015b} and Eq.~(\ref{eilenberger_ballistic}) can be integrated over the S width $d_S$. 

We work at temperatures close to the critical temperature and linearize the Eilenberger equation with respect to $\Delta/T_c$. In this approximation the Green's function takes the form:
\begin{equation}
\check g = 
\left(
\begin{array}{cc}
\hat g_N & \hat f \\
\hat {\tilde f} & \hat {\tilde g}_N
\end{array}
\right)_{\tau},
\label{gf_linearized}    
\end{equation}
where all the components are $4 \times 4$ matrices in the direct product of spin and sublattice spaces. The diagonal components $\hat g_N$ and $\hat {\tilde g}_N$  are to be calculated in the normal state of the superconductor, what is done in the previous section. The anomalous components $\hat f$ and $\hat {\tilde f}$ are of the first order with respect to $\Delta/T_c$ and contain singlet and N\'eel triplet correlations.  The resulting equation for the anomalous Green's function $\hat f$ takes the form:
\begin{eqnarray}
\left\{ i \omega_m \rho_x - \bm h_{eff} \bm \sigma \rho_z \rho_x, \hat f \right\} + \left[ \mu \rho_x, \hat f \right] +  \Delta \left( \rho_x \hat {\tilde g}_N - \hat g_N \rho_x \right) &- \nonumber \\ \frac{i}{\tau} 
\left( [\rho_+ \hat g_N \rho_+ + \rho_- \hat g_N \rho_-]\hat f 
- \hat f [\rho_+ \hat {\tilde g}_N \rho_+ + \rho_- \hat {\tilde g}_N \rho_-] \right) &- \nonumber \\
\frac{i}{\tau} 
\left( [\rho_+ \hat f \rho_+ + \rho_- \hat f \rho_-]\hat {\tilde g}_N 
- \hat g_N [\rho_+ \hat f \rho_+ + \rho_- \hat f \rho_-] \right)&= 0,
\label{f_linearized_eq}    
\end{eqnarray}
where $h_{eff} = h a_x/d_S$ is the absolute value of the effective staggered exchange field, induced in the superconductor due to proximity with the AF insulator. Now let $\bm h_{eff} = h_{eff}\bm e_z$. Then the solution of Eq.~(\ref{f_linearized_eq}) takes the form:
\begin{eqnarray}
\hat f = f_s \rho_x + f_t \rho_y \sigma_z , 
\label{f_sol}    
\end{eqnarray}
where 
\begin{align}
f_s = & \Delta \frac{-i\left(A^2-\tilde A^2\right)+\tau\left[(\tilde A-A)\mu + i h_{eff}(B+\tilde B)\right]}{2\left[(B+\tilde B)h_{eff} + \omega_m (A+\tilde A - i \mu \tau)\right]} ,~~~~ \label{fs_sol} \\
f_t = & \Delta \frac{i(B+\tilde B)(A - \tilde A + \omega_m \tau)}{2\left[(B+\tilde B)h_{eff} + \omega_m (A+\tilde A - i \mu \tau)\right]} . ~~~~
\label{ft_sol}    
\end{align}
In the clean limit, Eqs.~(\ref{fs_sol}) and (\ref{ft_sol}) are reduced to
	\begin{align}
	f_s = & \frac{\Delta}{2 i \omega_m} \Bigl( \frac{\omega_m - i \mu}{\sqrt{h_{eff}^2 + (\omega_m - i \mu)^2}}+\frac{\omega_m + i \mu}{\sqrt{h_{eff}^2 + (\omega_m + i \mu)^2}} \Bigr) \nonumber \\ 
	& + \frac{\Delta h_{eff}^2}{2 \mu \omega_m} \Bigl( \frac{1}{\sqrt{h_{eff}^2 + (\omega_m - i \mu)^2}}-\frac{1}{\sqrt{h_{eff}^2 + (\omega_m + i \mu)^2}} \Bigr), \label{fs_clean} \\
	f_t  = & \frac{h_{eff}\Delta}{2 \mu} \Bigl( \frac{1}{\sqrt{h_{eff}^2 + (\omega_m - i \mu)^2}}  -\frac{1}{\sqrt{h_{eff}^2 + (\omega_m + i \mu)^2}} \Bigr) .
	\label{ft_clean}    
	\end{align}
From Eq.~(\ref{f_sol}), it is seen that the triplet part of the anomalous Green's function is opposite at two sublattices (due to the $\rho_y$ term). Thus, the quasiclassical approach adequately captures and demonstrates the N\'eel triplets previously seen in the numerical solution (Fig.~\ref{fig:trip}). 

\begin{figure}[tb]
	\begin{center}
		\includegraphics[width=85mm]{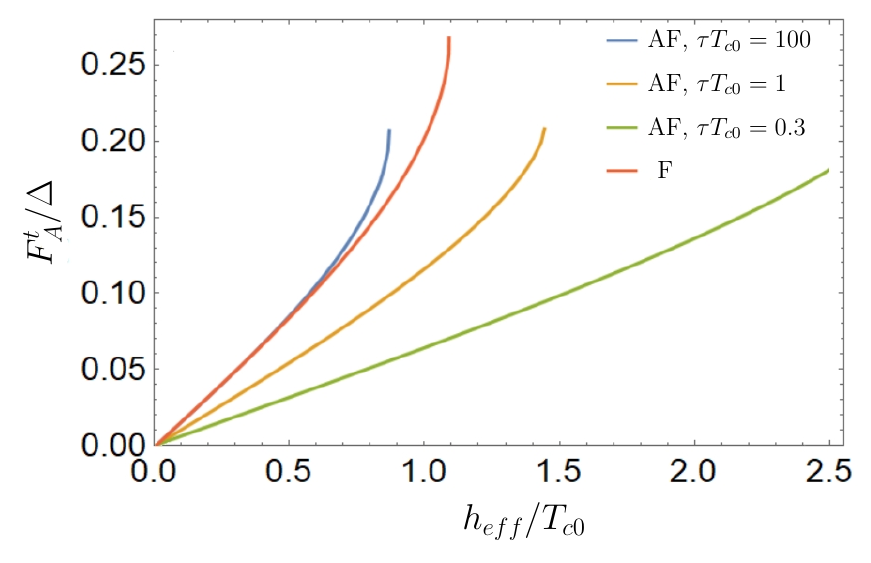}
		\caption{Anomalous Green's function of the N\'eel triplet correlations summed over positive Matsubara frequencies $F_A^t = \sum \limits_{\omega_m>0} f_t$ as a function of $h_{eff}$ for different values of the mean free time $\tau$.   Red line represents the same quantity for an S/F interface  with a ferromagnetic insulator producing the same value of the effective exchange field (but homogeneous, not staggered) in the superconductor. The S/F interface is not sensitive to impurities, for this reason only one line is shown for the ferromagnetic case. Each line ends at the critical value of $h_{eff}$ corresponding to the full suppression of superconductivity. We consider $\mu=0$ here.} \label{triplets_tau}
	\end{center}
\end{figure}

Eq.~(\ref{ft_clean}) demonstrates that the N\'eel triplets are odd-frequency correlations~\cite{Bergeret2005,Linder2019}. However, in contrast to the usual odd-frequency triplets in S/F heterostructures~\cite{Bergeret2005,Bergeret2018}, they are suppressed by impurities as demonstrated in Fig.~\ref{triplets_tau}, which presents the dependence of the N\'eel triplets on $h_{eff}$ at different impurity strengths. The suppression of the N\'eel triplets by impurities is because they are constituted by interband pairing and scattering from impurities mixes the states in the two bands. Furthermore, it is seen that in the absence of impurities the absolute value of the Neel triplet correlations is very close to the value of triplets at the S/F interface with the same parameters. The difference between them becomes essential at higher value of $h_{eff}$, where the suppression of superconductivity at the S/AF interface due to the antiferromagnetic gap also plays an important role.

\begin{figure}[tbh]
	\begin{center}
		\includegraphics[width=85mm]{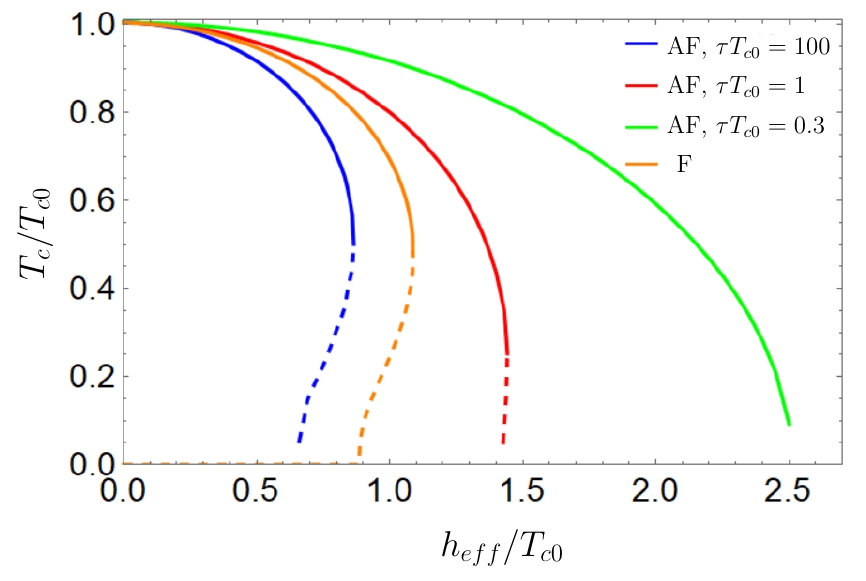}
		\caption{Critical temperature of the AF/S bilayer as a function of $h_{eff}$ for different values of mean free time $\tau$.  $\mu=0.01 T_{c0}$. Orange line represents $T_c(h_{eff})$ for an S/F interface with a ferromagnetic insulator producing the same value of the effective exchange field (but homogeneous, not staggered) in the superconductor. The dashed parts of the curves correspond to the unstable solutions, see text for further details.} \label{Fig:Tc_h_imp}
	\end{center}
\end{figure}

The critical temperature of the AF/S bilayer is calculated from the self-consistency equation
\begin{eqnarray}
1 = i \pi \lambda  T_c \sum \limits_{\omega_m} \frac{f_s}{\Delta} ,
\label{Tc}    
\end{eqnarray}
where $\lambda$ is the dimensionless coupling constant. The dependence of the critical temperature of the AF/S bilayer on the effective exchange field is presented in Fig.~\ref{Fig:Tc_h_imp} for different values of the impurity strength. We can observe two important physical phenomena in this figure.  

First, the critical temperature is indeed suppressed by the staggered exchange. In the clean case the efficiency of suppression by the staggered field is of the same order, and even higher, as the suppression by the ferromagnet with the same absolute value of the exchange field. The stronger suppression of the superconductivity by the staggered exchange as compared to the uniform ferromagnetic exchange field is explained by the presence of the antiferromagnetic gap at the Fermi surface [see Fig.~\ref{ldos_normal}], which prevents electronic states inside this gap from superconducting pairing. 

Second, the critical temperature for a given $h_{eff}$ grows with impurity concentration. That is, the impurities reduce the efficiency of suppression. It is in sharp contrast with the behavior of a F/S bilayer, where the degree of suppression is not sensitive to the impurity concentration. This highly unusual behavior results from  two facts, which act in parallel. First of all, as it is discussed above, the odd-frequency N\'eel triplets are suppressed by impurities. Second, the antiferromagnetic gap at the Fermi surface is also suppressed by impurities. It makes the corresponding electronic states available for superconducting pairing.

The dashed parts of the curves in Fig.~\ref{Fig:Tc_h_imp} represent the unstable branches of $T_c(h_{eff})$. It really means that starting from some value of $h_{eff}$ the self-consistency equation for the superconducting order parameter (beyond the linearized limit) has two different solutions corresponding to two different critical temperatures. One of them (with higher critical temperature) represents the stable solution and the other (with lower critical temperature) corresponds to the maximum of the system free energy, that is absolutely unstable. The situation is similar to the physics of S/F bilayers~\cite{Bobkova2014,Chandrasekhar1962,Clogston1962,Bergeret2018}.

\begin{figure}[tb]
	\begin{center}
		\includegraphics[width=85mm]{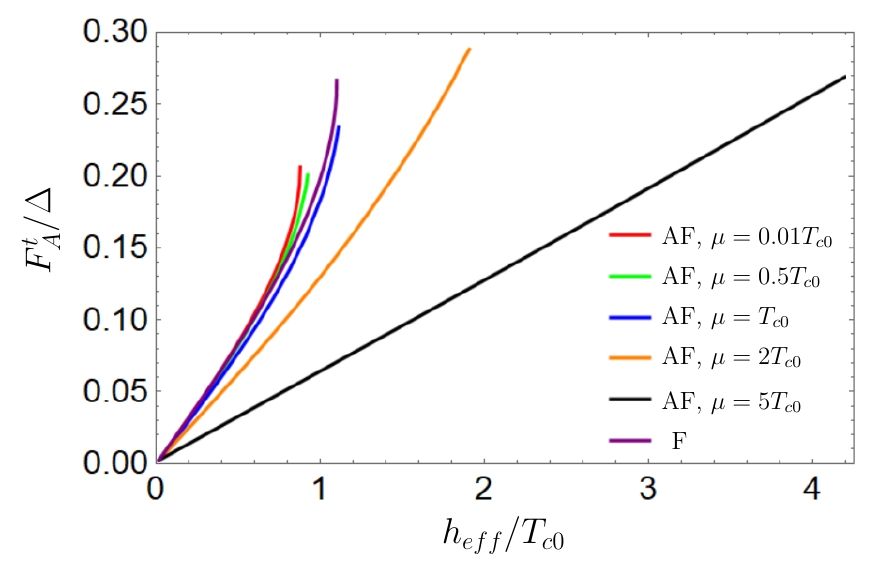}
		\caption{Anomalous Green's function of the Neel triplet correlations as a function of $h_{eff}$ for different values of  $\mu$. Each line ends at the critical value of $h_{eff}$ corresponding to the full suppression of superconductivity. We consider the clean limit $\tau^{-1}=0$ here.} \label{triplets_mu}
	\end{center}
\end{figure}

\begin{figure}[tb]
	\begin{center}
		\includegraphics[width=85mm]{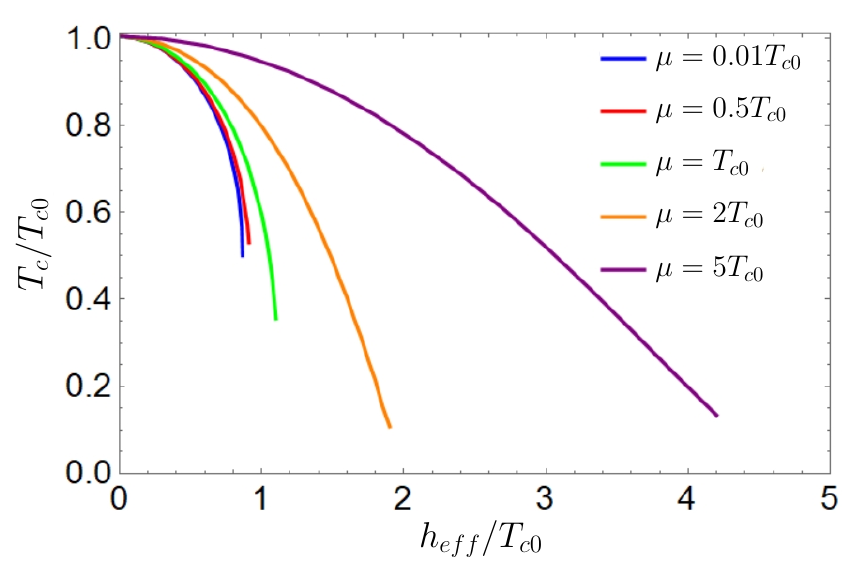}
		\caption{Critical temperature of the AF/S bilayer as a function of $h_{eff}$ for different values of $\mu$. Unstable branches are not shown. The results correspond to the clean limit $\tau^{-1}=0$.} \label{Fig:Tc_mu}
	\end{center}
\end{figure}

N\'eel triplets are also suppressed by finite values of $\mu$ for a given $h_{eff}$. This is demonstrated in Fig.~\ref{triplets_mu}. The physical reason for the suppression is understood from the discussion in Sec.~\ref{sec:relate} and the corresponding Fig.~\ref{fig:pairing}. Electrons can effectively pair if their energies are within the bulk superconducting gap $\Delta_0$ from the Fermi surface. N\'eel-type  or interband pairing involves electrons with different energies with the energy difference $2\mu$ between them (see Fig.~\ref{fig:pairing}). When this energy difference exceeds $\Delta_0$, the N\'eel-type pairing is not effective. This simple picture is valid for small values of $h_{eff}$ when the antiferromagnetic gap in the electron dispersion can be ignored. As it is seen from Fig.~\ref{triplets_mu}, at $h_{eff} \gtrsim \mu$ the suppression of  N\'eel triplets by $\mu$ is greatly weakened. The suppression  of the Neel triplets by nonzero $\mu$ naturally results in the reduced sensitivity of the critical temperature to the exchange field at larger $\mu$, as has been demonstrated in Fig.~\ref{Fig:Tc_mu}.

\section{Discussion}

A theoretical accounting of the bandgap created in the normal-state electron dispersion by the staggered spin-splitting induced by an adjacent N\'eel ordered AF has previously been considered~\cite{Krivoruchko1996}, without disorder. Furthermore, emergence of a finite spin-splitting in the S due to the commonly found uncompensated interface and spin-flip scattering offer other known mechanisms of proximity effect in an AF/S bilayer~\cite{Kamra2018}. Our analysis in this article has revealed a previously unexplored mechanism - converting spin-singlet into N\'eel spin-triplet Cooper pairs - of suppressing superconductivity in the S by an adjacent compensated AF. Furthermore, employing the two-sublattice quasiclassical Green's functions theory developed, our analysis reveals the effect of disorder on all these mechanisms. This seemingly complete picture allows us to envisage the various competing effects taking place in a realistic AF/S bilayer~\cite{Bell2003,Hubener2002,Seeger2021,Wu2013}.

Let us understand the qualitative dependencies of these mechanisms. The role of finite spin-splitting in S due to an uncompensated interface with an AF together with spin-flip scattering due to interfacial disorder is nearly identical to that of proximity effect due to an adjacent ferromagnet~\cite{Abrikosov1961,Kamra2018}. For the case of a compensated interface that we have considered here, both the bandgap opening mechanism and the induction of N\'eel triplets are most active when the Fermi wavevector in the normal state of the S (without AF) is close to the reciprocal space unit vector of the AF. In that case, the bandgap opening tends to make the normal-state of S into an insulator thereby diminishing density of states and weakening superconductivity. At the same time, under those conditions, the two bands (Fig.~\ref{fig:pairing}) are energetically close resulting in conversion of spin-singlet into N\'eel (or interband) spin-triplet Cooper pairs. Notably, disorder suppresses both these effects. Furthermore, if the S normal state does not have the required Fermi surface, both of these mechanisms become inactive and one recovers the Werthammer treatment of the proximity effect in the AF/S bilayer~\cite{Werthammer1966}, where the AF, if metallic, merely acts as a normal metal. 

Hence, to some extent, the experimental observation of a finite proximity effect and reduction of superconductivity in AF/S bilayers can be understood as due to the uncompensated interface that causes spin-splitting and spin-flip scattering. However, a proximity effect that turns out to be stronger than that due to a ferromagnet and that shows a maximum at specific AF thickness cannot be explained in this manner~\cite{Bell2003,Hubener2002,Wu2013}. These observations hint at the role of bandgap opening and/or the N\'eel triplets mechanism investigated here. A decrease in the average disorder with increasing thickness of the AF should allow for a stronger suppression of the superconductivity as per these two mechanisms. With further increase in the AF thickness, the S Cooper pairs do not leak into the metallic AF far enough and the proximity effect gets saturated. Furthermore, with reduced disorder, the weakening of superconductivity due to spin-flip scattering is decreased that may result in a recovery of the S critical temperature. Nevertheless, a fully detailed analysis of such a system should adequately consider a metallic AF, in contrast with our consideration of proximity effect due to an insulating AF.

\section{Summary}

We have demonstrated theoretically the emergence of unconventional spin-triplet Cooper pairs in a superconductor exposed via a compensated interface to an antiferromagnetic insulator. The pairing amplitude for these alternates sign on the lattice length scale, similar to the N\'eel spin order in the antiferromagnet. Thus, we call these N\'eel triplets and show that these are formed from pairing between electronic states from two different bands i.e., interband pairing. Formulating a two-sublattice quasiclassical Green's functions description of the bilayer, we investigate the effect of this interband pairing on the superconducting critical temperature. The ensuing powerful formulation provides a semi-analytic description and a clear physical understanding of the key phenomena as a function of disorder and chemical potential. Such N\'eel or interband triplets may already be playing an important role in experiments on proximity effect in superconductor/antiferromagnet hybrids.

\section*{Acknowledgments}
We thank Lina Johnsen for valuable discussions. We acknowledge financial support from the Spanish Ministry for Science and Innovation -- AEI Grant CEX2018-000805-M (through the ``Maria de Maeztu'' Programme for Units of Excellence in R\&D) and from the Russian Science Foundation via the RSF project No.22-22-00522.

\appendix

\section{Bogoliubov-de Gennes analysis: formulation and solution}

In this appendix, we formulate and solve the Bogoliubov-de Gennes equation~\cite{Zhu2016} which has been employed for the numerical analysis presented in Sec.~\ref{sec:BdG}. The system of our interest is described via the following Hamiltonian:
\begin{align}
H= - t \sum \limits_{\langle \bm{i}\bm{j}, \rangle \sigma} \hat c_{\bm{i} \sigma}^\dagger \hat c_{\bm{j} \sigma} + \sum \limits_{ \bm{i}} (\Delta_{\bm{i}} \hat c_{\bm{i}\uparrow}^\dagger \hat c_{\bm{i}\downarrow}^\dagger + H.c.) - \mu \sum \limits_{\bm{i} \sigma} \hat n_{\bm{i}\sigma} + \sum \limits_{\bm{i},\alpha \beta} \hat c_{\bm{i}\alpha}^\dagger (\bm{m}_{\bm{i}} \bm{\sigma})_{\alpha \beta} \hat c_{\bm{i}\beta} ,
\label{ham}
\end{align}
where $\bm i = (i_x,i_y)^T$ is the radius vector of the site and greek letters correspond to the spin indices. $\langle \bm i \bm j \rangle$ means summation over the nearest neighbors. $\Delta_{\bm i}$ and $\bm m_{\bm i}$ denote the on-site superconducting and magnetic order parameters at site $\bm i$, respectively. $\hat c_{\bm i \sigma}^\dagger (\hat c_{\bm i \sigma})$ creates (annihilates) an electron of spin $\sigma = \uparrow,\downarrow$ on the site $\bm i$, $t$ denotes the nearest-neighbor hopping integral, $\mu$ is the filling factor. $\hat n_{\bm i \sigma} = \hat c_{\bm i \sigma}^\dagger \hat c_{\bm i \sigma}$ is the particle number operator at site $\bm i$. We also define the vector of Pauli matrices in spin space $\bm \sigma = (\sigma_x, \sigma_y, \sigma_z)^T$. We assume that the antiferromagnet is of G-type. Then the magnetic order parameter can be taken in the form $\bm m_{\bm i} =  (-1)^{i_x+i_y} \bm m$ inside the antifferomagnet.  $x$- and $y$-axes are taken normal to the AF/S interface and parallel to it, respectively. It is assumed that the antiferromagnetism is due to the localized electrons and the amplitude of the on-site magnetization is not influenced by the adjacent metal. Therefore, we do not calculate $\bm m$ self-consistently and it is assumed to be constant inside the AF region. It has  been demonstrated for a similar AF/S/F structure~\cite{Johnsen2021}  that in the framework of the BdG approach the self-consistent calculation
of the antiferromagnetic order parameter  only gives a minor suppression of
the antiferromagnetic order parameter near the interface inside the antiferromagnet, and
does not lead to any qualitative changes inside the
superconductor, which is the focus of our study. We diagonalize the Hamiltonian (\ref{ham}) by the Bogolubov transformation:
\begin{align}
\hat c_{\bm i\sigma}=\sum\limits_n u^{\bm i}_{n\sigma}\hat b_n+v^{\bm i*}_{n\sigma}\hat b_n^\dagger , 
\label{bogolubov}
\end{align}
where $\hat b_n^\dagger (\hat b_n)$ are the creation (annihilation) operators of Bogolubov quasiparticles.
Then the resulting Bogolubov-de Gennes equations take the form:
	\begin{align}
	-\mu u^{\bm i}_{n,\sigma}-t\sum \limits_{\bm j \in \langle \bm i \rangle}u_{n,\sigma}^{\bm j} 
	+ \sigma \Delta_{\bm i} v^{\bm i}_{n,-\sigma}+(\bm {m}_{\bm i} \bm{\sigma})_{\sigma\alpha}u_{n,\alpha}^{\bm i} & = \varepsilon_n u_{n,\sigma}^{\bm i} \nonumber \\  
	-\mu v^{\bm i}_{n,\sigma}-t \sum \limits_{\bm j \in \langle \bm i \rangle}v_{n,\sigma}^{\bm j}
	+ \sigma \Delta_{\bm i}^* u^{\bm i}_{n,-\sigma}+(\bm{m}_{\bm i}\bm{\sigma}*)_{\sigma\alpha}v_{n,\alpha}^{\bm i} & = -\varepsilon_n v_{n,\sigma}^{\bm i}, 
	\label{bdg}
	\end{align}
where $\bm j \in \langle \bm i \rangle$ means summation over nearest neighbors of site $\bm i$ and $\varepsilon_n$ are the eigen-state energies of the Bogolubov quasiparticles. 

The superconducting order parameter in the S layer is calculated self-consistently:
\begin{align}
\Delta_{\boldsymbol{i}}= g\langle\hat c_{\bm{i} \downarrow} \hat c_{\bm{i} \uparrow} \rangle =  g \sum\limits_n (u_{n,\downarrow}^{\bm i} v_{n,\uparrow}^{\bm i*}(1-f_n)+u_{n,\uparrow}^{\bm i} v_{n,\downarrow}^{\bm i*}f_n),
\end{align}
where $g$ is the coupling constant. The quasiparticle distribution function is assumed to be the equilibrium Fermi distribution $f_n = \langle b_n^\dagger b_n \rangle = 1/(1+e^{\varepsilon_n/T})$.

Further we investigate the structure of superconducting  correlations  at the AF/S interface with infinite superconducting layer. The anomalous Green's function in Matsubara representation can be calculated as $F_{\bm i, \alpha \beta} = - \langle \hat c_{\bm i \alpha}(\tau) \hat c_{\bm i \beta}(0) \rangle$, where $\tau$ is the imaginary time. The component of this anomalous Green's function for a given Matsubara frequency $\omega_m = \pi T(2m+1)$ is calculated as follows: 
\begin{align}
F_{\bm i,\alpha\beta}(\omega_m)= \sum\limits_n (\frac{  u_{n,\alpha}^{\bm i} v_{n,\beta}^{\bm i*}}{i \omega_m -\varepsilon_n}+\frac{ u_{n,\beta}^{\bm i} v_{n,\alpha}^{\bm i*}}{i \omega_m +\varepsilon_n})
\end{align}
Only off-diagonal in spin space components, corresponding to opposite-spin pairs, are nonzero for the case under consideration. The singlet (triplet) correlations are described by $F_{\bm i}^{s,t}(\omega_m) = F_{\bm i,\uparrow \downarrow}(\omega_m) \mp F_{\bm i,\downarrow \uparrow}(\omega_m)$. Please note that the  on-site triplet correlations are odd in Matsubara frequency, as it should be according to the general fermionic symmetry. Therefore we calculate the sum over the positive Matsubara frequencies $F_{\bm i}^t = \sum \limits_{\omega_m>0} F_{\bm i}^t(\omega_m) $. 

\bibliography{Neel_pairing}

\end{document}